\documentclass{article}
\usepackage[T1]{fontenc}
\usepackage{microtype}
\usepackage{graphicx}
\usepackage{subcaption}
\usepackage{booktabs}
\usepackage{hyperref}

\usepackage[preprint]{sqe-preprint}
\usepackage{amsmath}
\usepackage{amssymb}
\usepackage{mathtools}
\usepackage{algorithm}
\usepackage{algorithmic}
\usepackage[capitalize,noabbrev]{cleveref}
\usepackage{xcolor}

\graphicspath{{figures/}}
\newcommand{\sqe}{\textsc{SQE}}

\pptitlerunning{When Does Query Expansion Help Agent Memory?}
\AtBeginDocument{\hypersetup{pdfsubject={Information Retrieval}, pdfkeywords={retrieval, query expansion, agent memory}}}

\begin{document}

\twocolumn[
  \pptitle{When Does Query Expansion Help Agent Memory?\\
    A Multi-Seed Study of Selective Expansion and Gate Reliability}

  \ppsetsymbol{equal}{*}
  \begin{ppauthorlist}
    \ppauthor{Salomon Diei}{kt}
  \end{ppauthorlist}
  \ppaffiliation{kt}{School of Computer Science and Engineering, KOREATECH, Cheonan, Republic of Korea}
  \ppcorrespondingauthor{Salomon Diei}{salomon@koreatech.ac.kr}
  \ppkeywords{retrieval, query expansion, agent memory, software engineering agents, negative results}

  \vskip 0.3in
]

\printAffiliationsAndNotice{}

\begin{abstract}
Long-horizon software agents store execution traces and later retrieve them
with natural-language questions, so the query and the memory are written in
different ``languages''. LLM query expansion can bridge this gap, but it costs
generation on every query. We study \emph{Selective Query-Side Expansion}
(\sqe{}), which expands a query with hypothetical execution traces and
paraphrases only when the top-1 dense retrieval score is low, then fuses all
rankings with Reciprocal Rank Fusion. Across 8 resampled memory stores with
rebuilt indices (4,000 queries over SWE-smith agent steps), \sqe{} reaches
69.4\% Recall@5 while expanding 48\% of queries; in a token-measured rerun it
uses 55\% fewer LLM tokens than always-on expansion. The observed gain over
dense retrieval is 0.9 Recall@5 points (95\% CI $[+0.1, +1.7]$, Holm-adjusted
$p=0.092$). Recall@1 drops, and the fixed top-1 gate shows no reliable gain
over random selection at a similar budget.
In an oracle analysis, expansion helps only 5.3\% of queries and hurts
4.2\%, so a perfect gate for this policy gains at most 5.3 points. We release
the memory stores, per-method summaries, and analysis scripts.
\end{abstract}

\section{Introduction}
\label{sec:intro}

Software-engineering agents accumulate long trajectories of observations,
commands, file edits, tests, and error traces. These records are useful only
if a later agent can retrieve the relevant fragment when it faces a related
problem. Retrieval is hard here because queries and memories are written in
different registers: a stored memory may be a traceback or a patch diff, while
the later query asks, for example, how an off-by-one line-number bug was fixed.

Query expansion is one remedy. HyDE~\cite{gao2023hyde} generates a
hypothetical document and retrieves with it; for agents, the analogous unit is a
hypothetical \emph{execution trace}. Expansion is expensive, though: each
expanded query can require several LLM calls. A useful gate should predict
which queries will benefit enough to justify the extra cost. This raises the
question we study:

\begin{center}
\emph{Can a cheap confidence signal decide which agent-memory\\queries are worth expanding?}
\end{center}

\begin{figure*}[t]
  \centering
  \includegraphics[width=0.92\textwidth]{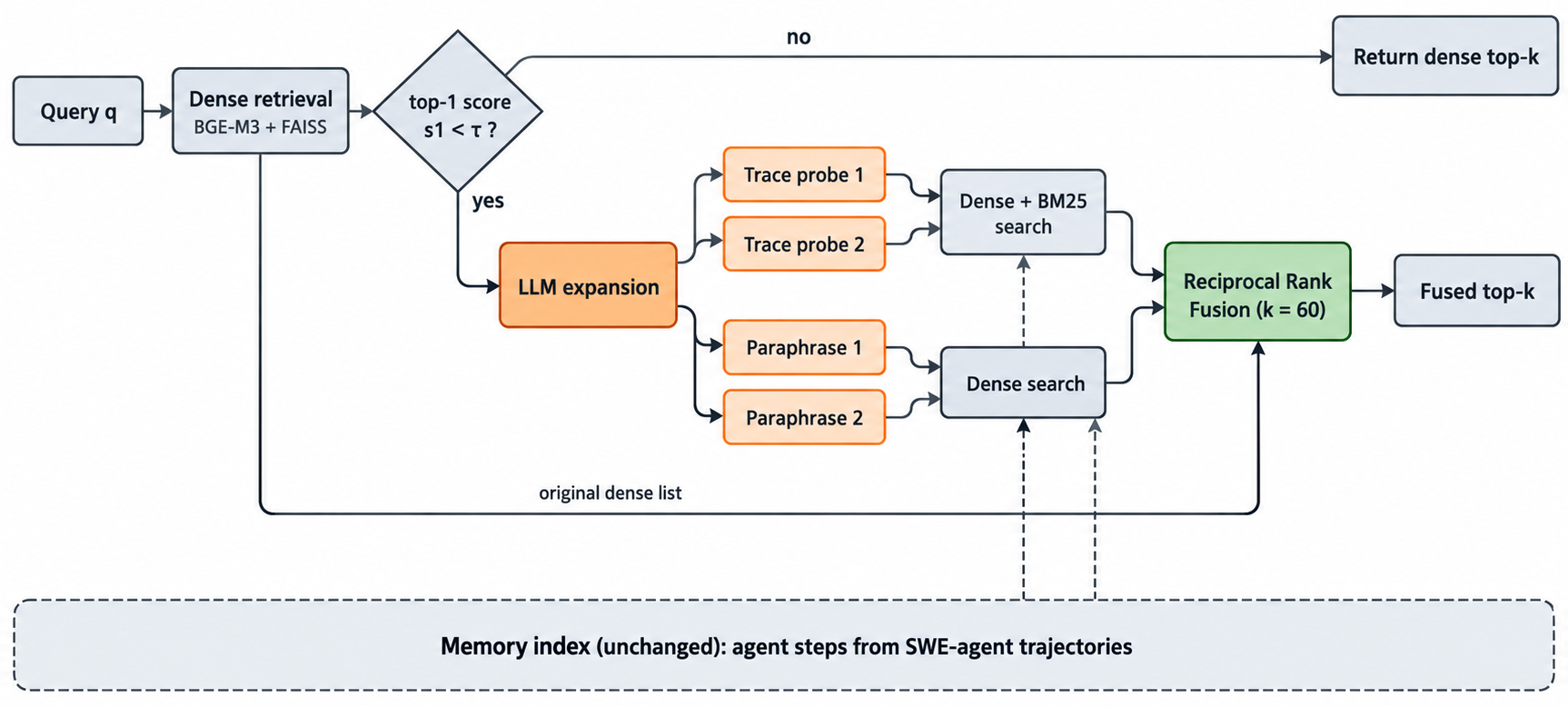}
  \caption{\sqe{} retrieval flow. Dense retrieval runs first; confident queries
  return the dense list unchanged. Queries whose top-1 score falls below $\tau$
  are expanded into two hypothetical trace probes (searched with dense and BM25)
  and two paraphrases (searched with dense only), all against the same unchanged
  index. The original dense list and the six probe lists are fused with RRF.}
  \label{fig:overview}
\end{figure*}

We implement \sqe{} (\cref{fig:overview}), which uses the top-1 dense score as
that signal, and evaluate it with an executed random-gate control at a similar
budget, eight resampled memory stores with rebuilt indices, measured token costs, and
an oracle-gate upper bound. Most of our findings are negative.

\paragraph{Contributions.}
\begin{enumerate}
  \item We formulate the trace--query mismatch in long-horizon software-agent
  memory and a selective, query-side-only expansion method that leaves the
  memory store and indices untouched (\cref{sec:method}).
  \item We run a controlled multi-seed study with an \emph{executed}
  random-gating baseline at a similar expansion rate and provider-reported token
  costs. The control helps assess whether the score adds value beyond the
  expansion budget itself (\cref{sec:results}).
  \item We compare the fixed top-1 gate with random selection and use an
  oracle to bound gains from gating this expansion policy (\cref{sec:gate}).
  \item We provide per-seed summaries, paired tests, and analysis scripts
  (\cref{sec:repro}).
\end{enumerate}

\section{Related Work}
\label{sec:related}

\paragraph{Dense and sparse retrieval.}
Dense passage retrieval maps queries and documents to a shared embedding
space~\cite{karpukhin2020dpr}; BM25 remains a strong lexical
baseline~\cite{robertson2009bm25}. Reciprocal Rank Fusion combines ranked
lists without training a ranker~\cite{cormack2009rrf}. We keep both indices
fixed and change only the query side.

\paragraph{Query expansion with generated text.}
HyDE embeds a generated hypothetical document for zero-shot dense
retrieval~\cite{gao2023hyde}, and Query2doc appends a generated pseudo-document
to the query~\cite{wang2023query2doc}. EAR expands and reranks candidate
queries~\cite{chuang2023ear}, and HyQE ranks contexts with hypothetical query
embeddings~\cite{zhou2024hyqe}. Later analysis shows that hypothetical-document
gains can partly reflect knowledge leakage from the generator~\cite{yoon2025hypothetical}.
We compare trace expansion with Query2doc and RM3-style pseudo-relevance
feedback~\cite{abduljaleel2004umass}.

\paragraph{Selective expansion and query performance prediction.}
Deciding per query whether to expand is an old idea. Query clarity predicts
retrieval quality without relevance labels~\cite{cronentownsend2002predicting},
and was used to apply pseudo-relevance feedback only where it is expected to
help~\cite{cronentownsend2004selective}. \citet{amati2004difficulty} found that
query-difficulty predictors make expansion more robust when applied
selectively. \sqe{} gates on the raw top-1 cosine score. We also test
predictors that use multiple retrieval scores, including a dense adaptation
of score magnitude and variance (SMV)~\cite{tao2014smv}. Adaptive retrieval
for LLMs makes a similar decision: Self-RAG learns when to
retrieve~\cite{asai2024selfrag}, Adaptive-RAG routes questions by predicted
complexity~\cite{jeong2024adaptiverag}, and FLARE retrieves when generation
confidence drops~\cite{jiang2023flare}. These methods gate retrieval itself;
\sqe{} always retrieves and gates only the query rewriting.

\paragraph{Agent memory and software agents.}
LoCoMo and LongMemEval evaluate long-term conversational memory of LLM
agents and assistants~\cite{maharana2024locomo,wu2025longmemeval}. AMA-Bench
and MemoryAgentBench evaluate memory over long-horizon and incremental
multi-turn interaction~\cite{zhao2026amabench,hu2025memae}. SWE-bench measures
whether an agent resolves real issues~\cite{jimenez2024swebench}, and SWE-smith
provides large-scale agent trajectories~\cite{yang2025swesmith}. Our benchmark
is narrower: retrieval of a known target episode, which we treat as an
intermediate signal; it does not measure task success.

\section{Method}
\label{sec:method}

Let $M=\{m_i\}_{i=1}^{N}$ be a store of execution traces. For a query $q$,
dense retrieval returns a ranked list $R_d(q)$ with top-1 cosine score
$s_1(q)$. \sqe{} expands $q$ iff $s_1(q)<\tau$; otherwise it returns $R_d(q)$
unchanged. An expanded query yields $K$ hypothetical traces and $P$ paraphrases
from an LLM. For each trace we retrieve a dense list and a BM25 list. For each paraphrase
we retrieve a dense list. We fuse the original dense list and the $2K+P$ probe
lists, using the top 20 entries of each list. Fusion assigns each retrieved
episode a score
\begin{equation}
  \mathrm{score}(m)=\sum_{j:\,m\in R_j}
       \frac{1}{c+\mathrm{rank}_j(m)},
\end{equation}
where $R_j$ is the $j$th ranked list. We use $\tau=0.65$ for the fixed gate,
$K=P=2$, and $c=60$. We return the top 10 episodes, and the generated traces
serve as retrieval probes.
\Cref{alg:sqe} summarizes the procedure.

\begin{algorithm}[t]
  \caption{Selective Query-Side Expansion}
  \label{alg:sqe}
  \begin{algorithmic}
    \STATE {\bfseries Input:} query $q$, threshold $\tau$, LLM $G$, budgets $K,P$
    \STATE $R_d, s_1 \leftarrow \textsc{Dense}(q)$
    \IF{$s_1 \ge \tau$}
      \STATE {\bfseries return} first 10 elements of $R_d$
    \ENDIF
    \STATE $\mathcal{T} \leftarrow G_{\text{trace}}(q,K)$, $\;\mathcal{P} \leftarrow G_{\text{para}}(q,P)$
    \STATE $\mathcal{L} \leftarrow \{R_d\} \cup \{\textsc{Dense}(p) : p \in \mathcal{P}\}$
    \STATE $\mathcal{L} \leftarrow \mathcal{L} \cup \{\textsc{Dense}(t), \textsc{BM25}(t) : t \in \mathcal{T}\}$
    \STATE {\bfseries return} first 10 elements of $\textsc{RRF}(\mathcal{L}, c)$
  \end{algorithmic}
\end{algorithm}

\section{Experimental Setup}
\label{sec:setup}

\paragraph{Data.} We use the \texttt{tool} split of the SWE-smith
trajectories~\cite{yang2025swesmith}, produced with SWE-agent and Claude~3.7
Sonnet. We split each trajectory into individual thoughts, actions, and
observations and retain formatted steps longer than 100 characters. A store holds
5,000 such steps. For each seed (42--49) we shuffle a shared pool of about
16,500 steps, take 5,000 as the store, rebuild both indices, and sample 500
evaluation targets from the store. The stores are therefore resampled, not
disjoint: two stores share about 30\% of their steps, and the 4,000 targets
contain 3,577 distinct episode IDs and 3,527 distinct target texts. For each
target, an LLM that sees the step writes
one question meant to identify it, and is told not to copy identifiers, error
codes, or variable names. Exactly one stored step counts as relevant per query,
so sibling steps from the same trajectory count as misses.

\paragraph{Models.} Dense retrieval uses BAAI/bge-m3 with FAISS. We use
BM25 with lowercase whitespace tokenization for sparse retrieval.
Qwen3.6-35B-A3B, served with vLLM, generates evaluation queries and
expansion probes.

\paragraph{Baselines.} Dense-Only, BM25-Only, Hybrid-RRF (dense + BM25),
Paraphrases-Only and HyDE-Traces-Only (always expand with one probe type),
Always-Expand (both probe types on every query), and
Random-Gated-Expansion, which runs the identical expansion code but
replaces the gate with seeded random decisions at a comparable rate. The random
gate is the main control: it separates the value of \emph{expanding some
queries} from the value of \emph{choosing which ones}.

Query2doc generates one pseudo-document per query with Qwen3.6-35B-A3B
and four examples from another seed. Examples exclude the current query's
source task, but their tasks can occur in the evaluated memory store.
Query2doc-Dense concatenates the query and pseudo-document, while
Query2doc-BM25 repeats the query five times before appending the document.
Our RM3-style baseline uses the top ten BM25 results as feedback documents
and selects up to ten expansion terms. It assigns half the query weight to
the feedback terms and uses the resulting weights for BM25 scoring.

\paragraph{Metrics and inference.} Recall@$k$ is the fraction of queries
whose target episode appears in the top $k$ results. We report it at
$k\in\{1,5,10\}$. We record provider-reported generation tokens and client
wall-clock retrieval time in a seed-42 rerun. We use two-sided paired
bootstrap tests over query rows with Holm correction across nine contrasts.
The 95\% intervals are unadjusted and pool queries across overlapping
stores, so we treat them as descriptive (\cref{sec:limits}).

\section{Results}
\label{sec:results}

\begin{table}[t]
  \caption{Retrieval over 8 resampled memory stores (mean $\pm$ std over
  seeds using sample standard deviation, \%). \sqe{} expands 48.2\% of queries on average; the random gate
  expands 45.9\%.}
  \label{tab:multiseed}
  \centering
  \small
  \resizebox{\columnwidth}{!}{\begin{tabular}{lrrrr}
\toprule
Method & Seeds & Recall@1 & Recall@5 & Recall@10 \\
\midrule
Dense-Only & 8 & 45.5 $\pm$ 3.1 & 68.5 $\pm$ 2.7 & 74.2 $\pm$ 2.5 \\
BM25-Only & 8 & 16.9 $\pm$ 0.7 & 30.1 $\pm$ 1.4 & 34.5 $\pm$ 1.4 \\
Hybrid-RRF & 8 & 32.5 $\pm$ 1.5 & 63.5 $\pm$ 2.5 & 71.8 $\pm$ 2.6 \\
Always-Expand & 8 & 39.8 $\pm$ 3.1 & 69.2 $\pm$ 3.4 & 75.9 $\pm$ 2.3 \\
Random-Gated-Expansion & 8 & 42.6 $\pm$ 2.9 & 68.9 $\pm$ 3.0 & 75.0 $\pm$ 2.6 \\
Selective-QE & 8 & 44.5 $\pm$ 3.4 & 69.4 $\pm$ 2.7 & 75.3 $\pm$ 2.2 \\
Query2doc-Dense & 8 & 37.9 $\pm$ 3.1 & 59.8 $\pm$ 3.9 & 66.3 $\pm$ 3.5 \\
Query2doc-BM25 & 8 & 20.2 $\pm$ 1.8 & 34.3 $\pm$ 1.5 & 40.0 $\pm$ 2.6 \\
RM3-BM25 & 8 & 7.8 $\pm$ 0.8 & 18.3 $\pm$ 1.6 & 22.4 $\pm$ 1.6 \\
\bottomrule
\end{tabular}
}
\end{table}

\begin{table}[t]
  \caption{Paired bootstrap of \sqe{} minus each baseline, Recall@5 points,
  pooled over 4,000 query rows. Intervals are unadjusted. Holm $p$ corrects
  the two-sided $p$-values for nine contrasts. Entries of 0.000 fall below
  0.001 at the reported precision.}
  \label{tab:paired}
  \centering
  \small
  \resizebox{\columnwidth}{!}{\begin{tabular}{lrrrrr}
\toprule
Baseline & Queries & $\Delta$R@5 & 95\% CI & $p$ & Holm $p$ \\
\midrule
Dense-Only & 4000 & +0.9 & [+0.1, +1.7] & 0.031 & 0.092 \\
Hybrid-RRF & 4000 & +5.9 & [+4.8, +7.0] & 0.000 & 0.000 \\
Always-Expand & 4000 & +0.1 & [-0.4, +0.7] & 0.630 & 0.630 \\
Random-Gated-Expansion & 4000 & +0.5 & [-0.2, +1.1] & 0.170 & 0.340 \\
Paraphrases-Only & 4000 & +1.1 & [+0.3, +1.9] & 0.010 & 0.042 \\
HyDE-Traces-Only & 4000 & +5.6 & [+4.7, +6.6] & 0.000 & 0.000 \\
Query2doc-Dense & 4000 & +9.6 & [+8.3, +10.9] & 0.000 & 0.000 \\
Query2doc-BM25 & 4000 & +35.1 & [+33.4, +36.8] & 0.000 & 0.000 \\
RM3-BM25 & 4000 & +51.1 & [+49.4, +52.8] & 0.000 & 0.000 \\
\bottomrule
\end{tabular}
}
\end{table}

\begin{figure}[t]
  \centering
  \includegraphics[width=\columnwidth]{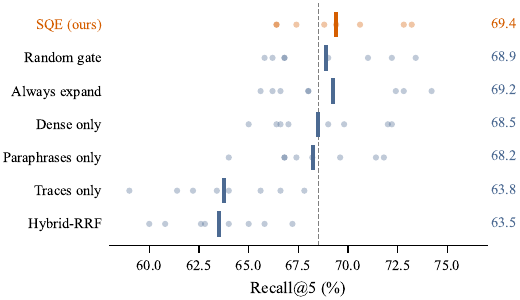}
  \caption{Recall@5 per seed (dots) and 8-seed mean (bar). The dashed line is
  the Dense-Only mean. This figure shows the original probe and gate ablations.}
  \label{fig:multiseed}
\end{figure}

\paragraph{\sqe{} is slightly better than dense retrieval at Recall@5, and worse at Recall@1.}
\Cref{tab:multiseed} shows \sqe{} at 69.4\% Recall@5 against 68.5\% for
Dense-Only. The paired difference is $+0.9$ points with 95\% CI $[+0.1,+1.7]$
(\cref{tab:paired}). At query level, \sqe{} recovers 140 targets that dense
retrieval missed but loses 105 that it found, a net gain of 35 out of 4,000
(\cref{tab:winloss}). Recall@1 falls from 45.5\% to 44.5\%.
Across the eight stores, \sqe{}'s Recall@5 has a sample standard deviation of
2.7 points (\cref{fig:multiseed}). The appendix reports the seed-42 ablations
and illustrative wins and losses (\cref{tab:seed42,tab:cases}).

\paragraph{Query2doc and RM3 trail dense retrieval.}
Query2doc-Dense reaches 59.8\% Recall@5, versus 68.5\% for Dense-Only.
Query2doc-BM25 reaches 34.3\%, a 4.1-point gain over BM25-Only.
RM3-BM25 reaches 18.3\%, an 11.9-point drop from BM25-Only. All three
expansion baselines trail \sqe{} at Recall@5 (\cref{tab:paired}).

\paragraph{The fixed top-1 gate shows no reliable gain over a random gate.}
Against Random-Gated-Expansion at a similar budget, \sqe{} gains $+0.5$
points with CI $[-0.2,+1.1]$. Against Always-Expand the difference is $+0.1$
$[-0.4,+0.7]$. The two policies achieve similar mean Recall@5, and their
query-level outcomes can differ.

\begin{figure}[t]
  \centering
  \includegraphics[width=\columnwidth]{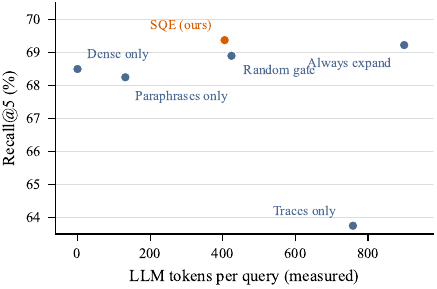}
  \caption{Cost against quality. $x$: provider-measured LLM tokens per query
  (seed 42); $y$: eight-seed mean Recall@5 from the archived retrieval
  runs. \sqe{} and the random gate have similar recall to Always-Expand
  and use roughly half the generation tokens.}
  \label{fig:cost}
\end{figure}

\paragraph{Selectivity is still a large cost saving.}
\Cref{fig:cost} pairs the eight-seed Recall@5 means with costs from the seed
42 token-measured rerun. \sqe{} uses 405 tokens and 2.8\,s per query versus
899 tokens and 5.4\,s for Always-Expand (\cref{tab:tokens}, appendix). The
archived recalls are 69.4\% and 69.2\%, respectively. Dense-Only uses no
LLM tokens and reaches 68.5\% Recall@5. HyDE-Traces-Only reaches 63.8\%
Recall@5 across seeds and uses 758 tokens per query in the rerun. The
Always-Expand run, which combines the trace and paraphrase probes, reaches
69.2\%. This comparison measures the tested configurations.

\section{What the Gate Can Predict}
\label{sec:gate}

\paragraph{Low confidence marks hard queries.}
On seed 42, expanded queries reach 61.7\% Recall@5 and non-expanded ones
75.9\% (\cref{tab:gatediag}, appendix), so the top-1 score does find harder
queries. This split cannot show whether expansion helps them, because each
group is measured only after its own treatment; the oracle analysis below
compares both treatments on the same queries.

\begin{figure}[t]
  \centering
  \includegraphics[width=\columnwidth]{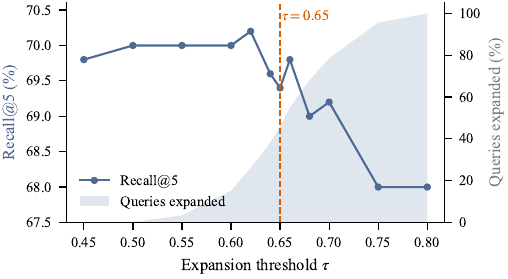}
  \caption{Post-hoc threshold sweep on seed 42. Recall@5 (line, left axis)
  varies by about 2 points while the fraction of expanded queries (shaded,
  right axis) spans 0--100\%.}
  \label{fig:threshold}
\end{figure}

\paragraph{The threshold barely matters.}
\Cref{fig:threshold} sweeps $\tau$ on seed 42. Recall@5 stays within a
2.2-point band while the expansion rate goes from 0\% to 100\%, and the best
post-hoc threshold (0.62) beats the fixed one by only 0.8 points. Held-out
selection does not change this picture: leave-one-seed-out threshold selection
gives $+0.6$ points over dense retrieval with CI $[-0.1,+1.3]$
(\cref{tab:loso}, appendix), and BM25-agreement gate variants do
not close the gap (\cref{tab:variants}, appendix).

\paragraph{Exploratory query-performance predictors.}
For each held-out seed, we tuned each of seven gate features on the other
seeds after removing rows whose source tasks occurred among the held-out
queries. Gate outcomes reuse the executed Dense-Only and Always-Expand
rankings. At a mean expansion rate of 32.7\%, the SMV gate gained 1.1
Recall@5 points over the expected recall of a random gate at the same budget,
with a positive difference in all eight seeds. The two-sided seedwise sign
test gives nominal $p=0.008$ and Bonferroni-adjusted $p=0.055$ across the
seven features. In a post-hoc sensitivity check, we excluded the 39 rows
whose recomputed dense top ten differed from the archived rankings. The
rounded gain remained 1.1 points, and all eight seed differences remained
positive. The overlapping stores limit inference. Confirming the SMV
result requires an evaluation on disjoint source tasks.

\begin{figure}[t]
  \centering
  \includegraphics[width=\columnwidth]{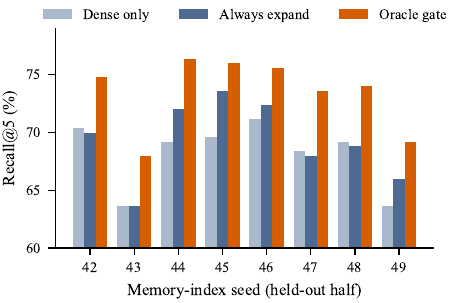}
  \caption{Held-out oracle headroom. Even a perfect per-query gate gains about
  5 points over dense retrieval; always expanding gains about 1.}
  \label{fig:headroom}
\end{figure}

\paragraph{The oracle bounds gains for this expansion policy.}
To bound what a better gate could achieve, we combine executed Dense-Only and
Always-Expand rows on the held-out half of each seed. An oracle that expands
exactly the queries that expansion helps would reach 73.5\% Recall@5, against
68.2\% for dense retrieval (\cref{fig:headroom}). Expansion helps 5.3\% of
queries and hurts 4.2\%; the remaining 90\% have the same outcome either way.
The rare changes keep the absolute gain small even under perfect selection.
The prediction target is whether expansion changes the outcome for a given query.

\section{Limitations}
\label{sec:limits}

\paragraph{Queries.} Evaluation queries are generated by an LLM that sees the
target trace, which can leak target-specific wording and overstate
discriminability. A human-audit packet is prepared but unlabeled.

\paragraph{Relevance labels.} Only the exact source step is relevant. Adjacent
steps of the same trajectory can answer the same question, so Recall@$k$ may
understate all methods, and possibly expansion more than dense retrieval.

\paragraph{No downstream outcome.} Recall@$k$ is an intermediate signal; we do
not measure whether retrieved memories change agent Pass@1 on SWE-bench-style
tasks~\cite{jimenez2024swebench}.

\paragraph{Statistics.} The stores share about 30\% of their episodes and
some evaluation targets. The pooled bootstrap intervals may understate
uncertainty. The Holm-adjusted comparison with Dense-Only gives $p=0.092$.
A confirmatory study should pre-register a primary contrast, hold out
source tasks, and account for clustering.

\paragraph{Scope.} One embedder, one generator, one trace source, and one
memory size. The original runs used an unpinned upstream dataset revision
and did not record generator inference seeds. The released sample and
rankings preserve the evaluated inputs and outcomes.

\section{Reproducibility}
\label{sec:repro}

Code, paper sources, and per-seed summaries are at
\url{https://github.com/Salomondiei08/sqe-experiment}; memory stores,
evaluation pairs, and result artifacts are at
\url{https://huggingface.co/datasets/TheReinventGuy/sqe-experiment}.
\texttt{scripts/07\_verify\_experiment.py} recomputes all metrics from result
files for the original seed-42 experiment. Scripts 09 and 11 regenerate
the multi-seed tables, and script 48 reproduces the exploratory gate analysis.
The repository's \texttt{make\_figures.py} regenerates the result figures
from the released JSON. \Cref{fig:overview} illustrates the method.

\section{Conclusion}
\label{sec:conclusion}

Selective query-side expansion halves the generation cost of always-on
expansion for agent-memory retrieval with similar Recall@5. The fixed
top-1 gate shows no reliable gain over random selection, and SQE improves on
dense retrieval by under one point. The exploratory SMV result suggests a
small gain from a richer score feature. The oracle bounds the gains available
from choosing when to apply this expansion policy. Future work should
evaluate learned gates against matched random budgets and measure downstream
agent success.

\section*{Impact Statement}
This work evaluates retrieval of archived software-engineering agent steps.
Evaluation queries and expansion probes are generated from or used with
public trajectories. We measure retrieval outcomes and generation costs.
The effect on agents' downstream behavior remains open.

\section*{Acknowledgments and Disclosure}
An earlier version of this work was presented at NAFSIK 2026. The author used
LLM-based assistance to organize the experiment package, write audit scripts,
edit text, and draw \cref{fig:overview}. The retrieval metrics come from
archived experiment outputs.

\bibliography{references}
\bibliographystyle{sqe-preprint}

\newpage
\appendix
\onecolumn

\section{Additional Results}

\begin{table}[h]
  \caption{Seed-42 results for the original retrieval and probe ablations
  (\%). See \cref{tab:multiseed} for the additional expansion baselines.}
  \label{tab:seed42}
  \centering
  \small
  \begin{tabular}{lrrrrr}
\toprule
Method & Recall@1 & Recall@5 & Recall@10 & MRR & Expansion \%
\\
\midrule
Dense-Only & 49.0 & 69.8 & 75.4 & 57.4 & -- \\
BM25-Only & 17.6 & 31.4 & 35.8 & 23.3 & -- \\
Hybrid-RRF & 35.4 & 64.0 & 73.2 & 47.4 & -- \\
Paraphrases-Only & 47.0 & 68.2 & 75.0 & 56.0 & 100.0 \\
HyDE-Traces-Only & 36.2 & 64.0 & 70.4 & 47.9 & 100.0 \\
Always-Expand & 42.2 & 68.0 & 75.6 & 53.2 & 100.0 \\
Random-Gated-Expansion & 46.0 & 69.0 & 75.8 & 55.3 & 47.0 \\
Selective-QE & 47.2 & 69.4 & 75.4 & 56.4 & 46.0 \\
\bottomrule
\end{tabular}

\end{table}

On seed 42, \sqe{} reaches 69.4\% Recall@5 against 69.8\% for Dense-Only
($-0.4$) and 64.0\% for Hybrid-RRF ($+5.4$); Recall@1 is 47.2\% against 49.0\%.
The multi-seed aggregate in \cref{tab:multiseed} is the primary evidence.

\begin{table}[h]
  \caption{Query-level top-5 wins and losses of \sqe{} against Dense-Only.}
  \label{tab:winloss}
  \centering
  \small
  \begin{tabular}{lrrrrrr}
\toprule
Split & N & Both hit & SQE win & SQE loss & Both miss & Net \\
\midrule
seed 42 & 500 & 334 & 13 & 15 & 138 & -2 \\
seed 43 & 500 & 314 & 18 & 18 & 150 & 0 \\
seed 44 & 500 & 347 & 17 & 13 & 123 & 4 \\
seed 45 & 500 & 335 & 18 & 10 & 137 & 8 \\
seed 46 & 500 & 350 & 16 & 11 & 123 & 5 \\
seed 47 & 500 & 320 & 17 & 15 & 148 & 2 \\
seed 48 & 500 & 322 & 22 & 11 & 145 & 11 \\
seed 49 & 500 & 313 & 19 & 12 & 156 & 7 \\
\midrule
All & 4000 & 2635 & 140 & 105 & 1120 & 35 \\
\bottomrule
\end{tabular}

\end{table}

\begin{table}[h]
  \caption{Provider-reported generation tokens and client wall-clock
  retrieval time (seed 42, 500 queries, expansion cache disabled).}
  \label{tab:tokens}
  \centering
  \small
  \begin{tabular}{lrrrrr}
\toprule
Method & Expanded & Calls/q & Prompt tok/q & Total tok/q & Latency s/q \\
\midrule
Selective-QE & 46.0 & 1.38 & 209.2 & 405.1 & 2.77 \\
Always-Expand & 100.0 & 2.99 & 455.5 & 899.3 & 5.37 \\
HyDE-Traces-Only & 100.0 & 2.00 & 367.0 & 758.2 & 3.99 \\
Paraphrases-Only & 100.0 & 1.00 & 88.5 & 131.6 & 1.02 \\
Random-Gated-Expansion & 47.0 & 1.41 & 214.6 & 423.9 & 2.40 \\
\bottomrule
\end{tabular}

\end{table}

\begin{table}[h]
  \caption{Seed-42 gate split by expansion decision.}
  \label{tab:gatediag}
  \centering
  \small
  \begin{tabular}{lrrrr}
\toprule
Subset & Count & Hits@5 & Misses@5 & Recall@5
\\
\midrule
Expanded & 230 & 142 & 88 & 61.7 \\
Not expanded & 270 & 205 & 65 & 75.9 \\
\bottomrule
\end{tabular}

\end{table}

\begin{table}[h]
  \caption{Leave-one-seed-out top-1 threshold selection.}
  \label{tab:loso}
  \centering
  \small
  \begin{tabular}{lrrrrrr}
\toprule
Test seed & Train seeds & Threshold & Gate R@5 & Dense R@5 & Delta & Expansion \\
\midrule
42 & 43,44,45,46,47,48,49 & 0.64 & 69.6 & 69.8 & -0.2 & 38.2 \\
43 & 42,44,45,46,47,48,49 & 0.62 & 66.6 & 66.4 & +0.2 & 27.6 \\
44 & 42,43,45,46,47,48,49 & 0.62 & 72.6 & 72.0 & +0.6 & 28.8 \\
45 & 42,43,44,46,47,48,49 & 0.62 & 69.8 & 69.0 & +0.8 & 28.8 \\
46 & 42,43,44,45,47,48,49 & 0.62 & 72.0 & 72.2 & -0.2 & 29.0 \\
47 & 42,43,44,45,46,48,49 & 0.64 & 67.4 & 67.0 & +0.4 & 42.6 \\
48 & 42,43,44,45,46,47,49 & 0.66 & 68.6 & 66.6 & +2.0 & 58.8 \\
49 & 42,43,44,45,46,47,48 & 0.64 & 66.4 & 65.0 & +1.4 & 41.2 \\
\midrule
All & -- & -- & 69.1 & 68.5 & +0.6 & 36.9 \\
\bottomrule
\end{tabular}

\end{table}

\begin{table}[h]
  \caption{Held-out score-only and BM25-agreement gate variants.}
  \label{tab:variants}
  \centering
  \small
  \begin{tabular}{llrrrr}
\toprule
Seed & Policy & Threshold & Test R@5 & Dense Test R@5 & Expansion \\
\midrule
42 & score\_and\_no\_bm25\_agreement & 0.50 & 70.4 & 70.4 & 0.0 \\
43 & score\_or\_no\_bm25\_agreement & 0.65 & 64.0 & 63.6 & 79.6 \\
44 & score\_only & 0.75 & 72.0 & 69.2 & 96.4 \\
45 & score\_or\_no\_bm25\_agreement & 0.70 & 73.6 & 69.6 & 90.4 \\
46 & score\_only & 0.65 & 72.0 & 71.2 & 51.6 \\
47 & score\_and\_no\_bm25\_agreement & 0.45 & 68.4 & 68.4 & 0.0 \\
48 & score\_and\_no\_bm25\_agreement & 0.70 & 68.8 & 69.2 & 55.6 \\
49 & score\_only & 0.60 & 64.4 & 63.6 & 19.6 \\
\midrule
Mean & -- & -- & 69.2 & 68.2 & 49.1 \\
\bottomrule
\end{tabular}

\end{table}

\begin{table}[h]
  \caption{Seed-42 examples where \sqe{} and Dense-Only differ at top-5.
  Illustrative only.}
  \label{tab:cases}
  \centering
  \scriptsize
  \resizebox{\textwidth}{!}{\begin{tabular}{lllllp{0.45\linewidth}}
\toprule
Case & Query ID & Dense rank & SQE rank & Expanded & Query \\
\midrule
SQE gain & q\_0014 & >10 & 3 & yes & What regular expression pattern was used to define the digit group variable? \\
SQE gain & q\_0063 & 8 & 3 & yes & Did the test script confirm that the parser correctly handles NULL comparisons... \\
SQE loss & q\_0002 & 1 & 7 & yes & Did the test script confirm that the fix for the FOO value was successful? \\
SQE loss & q\_0031 & 1 & 7 & yes & What was the output when I ran the simple test script to verify the initial cal... \\
\bottomrule
\end{tabular}
}
\end{table}

\end{document}